\documentclass[preprint,12pt]{aastex}               
\usepackage{epsfig}

\shortauthors{D.~A.~Uzdensky}

\begin{document}

\title{Shear-Driven Field-Line Opening and the Loss of a Force-Free 
Magnetostatic Equilibrium}
\author{Dmitri A. Uzdensky}
\affil{Institute for Theoretical Physics, University of California} 
\affil{Santa Barbara, CA 93106}
\email{uzdensky@itp.ucsb.edu}
\date{\today}

\begin{abstract}
This paper discusses a quasi-static evolution of a force-free magnetic 
field under slow sheared footpoint motions on the plasma's boundary, an 
important problem with applications to the solar and accretion disk coronae. 
The main qualitative features of the evolution (such as field-line expansion 
and opening) are considered and a comparison is made between two different 
geometrical settings: the Cartesian case with translational 
symmetry along a straight line, and the axisymmetric case with axial 
symmetry around the rotation axis. The main question addressed in the 
paper is whether a continuous sequence of force-free equilibria describes 
the evolution at arbitrarily large values of the footpoint displacement or 
the sequence ends abruptly and the system exhibits a loss of equilibrium at 
a finite footpoint displacement. After a formal description of the problem, 
a review/discussion of the extensive previous work on the subject is given. 
After that, a series of simple scaling-type arguments, explaining the key 
essential reason for the main qualitative difference between the two 
geometry types, is presented. It is found that, in the Cartesian case, 
force-free equilibria exist at arbitrarily large values of shear and 
the field approaches the open state only at infinite shear, whereas 
in the axisymmetric case the field opens up already at a finite shear.
\end{abstract}



\section{Introduction and Geometrical Settings}
\label{sec-intro}

In studies of force-free coronal magnetic fields in solar physics, as 
well as in a closely related and essentially very similar problem of 
accretion disk magnetospheres, there has been some controversy regarding 
the issue of the {\it loss of equilibrium}. This controversy has arisen 
from the problem of finding a continuous sequence of force-free equilibria 
in the corona, invoked to represent a time evolution of the coronal magnetic
field under slow plasma motions in the Sun's photosphere. Indeed, the 
footpoints of the magnetic field lines are frozen into the photosphere 
and hence the photospheric motions lead to continuous shearing of the 
magnetic field. Under the assumption of ideal magnetohydrodynamics (MHD), 
if these motions are much slower than the Alfv{\'e}n velocity in the 
corona (an assumption justified by a very low plasma density in the 
corona), the coronal magnetic field progresses through a sequence of 
equilibria. An important aspect of the problem is that the domain under 
consideration is infinite, so that the field lines can expand freely into 
space, instead of being confined to a finite-size box.

When trying to build a theoretical model of this process, 
one typically starts with a potential (no shear) field, and
then gradually increases the shear. This initial potential 
field is taken to be closed, which means that both footpoints 
of each field line lie on the surface, i.e., no field line 
extends to infinity. The critical question then is whether 
one should be able to find a force-free equilibrium 
configuration (with the same topology as that of the original 
potential field) as the shear is increased indefinitely, or one 
should reach a certain critical point, beyond which no same-topology 
equilibrium solutions can be found (loss of equilibrium). 

Even though this question has first been tackled in the context of 
the solar corona, a very similar process has also been investigated 
in the context of accretion disk coronae, where the sheared footpoint 
motion arises naturally from the differential rotation of a Keplerian 
disk or from the relative star--disk rotation (see van~Ballegooijen 1994; 
Goodson et al. 1999; Uzdensky et al. 2002; Lovelace et al. 1995; 
Uzdensky 2002). 

\begin{figure}
\plotone{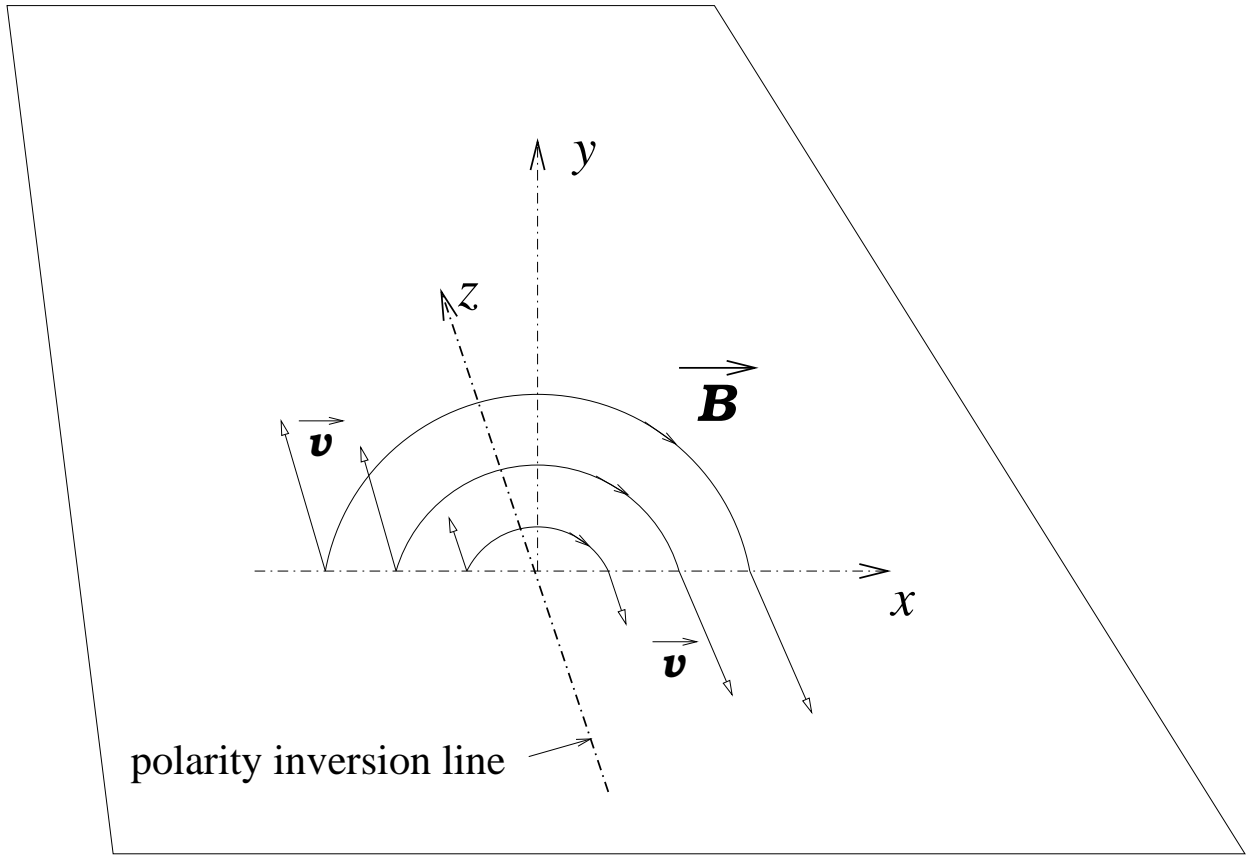}
\figcaption{Cartesian geometry with translational symmetry 
along the polarity inversion line (the $z$~axis); vector 
${\bf v}$ represents the footpoint displacement along this 
line. 
\label{fig-cartesian}}
\end{figure}

In either context, some simplifying assumptions are usually made 
in order to make the problem tractable. One of the most important 
is the assumption that there is one ignorable direction on the 
photospheric surface, i.e., a direction along which the fields 
are constant. There are two different symmetry classes that are 
most often studied:

1) {\it Cartesian (or plane) geometry} (see Fig.~\ref{fig-cartesian}),
the photosphere being an infinite plane and the corona --- a half-space
above this plane. The field line topology is that of a straight line dipole 
placed on or somewhere beneath the plane. The footpoints are displaced 
along the line dipole axis (also called the polarity inversion line) 
and the system possess translational symmetry along this axis. This
problem is usually studied in Cartesian coordinates, with the ignorable 
direction (the line dipole axis) denoted by, say, $z$ and the direction 
perpendicular to the plane by~$y$. There have been extensive analytical 
(Low 1977, 1982, 1990; Birn et al. 1978; Priest \& Milne 1980; Birn \&
Schindler 1981; Aly 1984, 1985, 1990, 1993, 1994; Priest \& Forbes 1990) 
and numerical studies of this problem. The latter can be subdivided further 
into numerical computations of sequences of force-free equilibria (Sturrock 
\& Woodbury 1967; Jockers 1978; Klimchuk et al. 1988; Klimchuk \& Sturrock 
1989; Finn \& Chen 1990; Wolfson \& Verma 1991) and full MHD numerical 
simulations (e.g., Biskamp \& Welter 1989; Amari et al. 1996a).

2) {\it axisymmetric geometry}, with axial (or cylindrical) symmetry 
around the $z$ axis ($\phi$-direction). This case is usually treated 
in either cylindrical ($\rho,z,\phi$) or spherical ($r,\theta,\phi$)
coordinates. The footpoints rotate in the azimuthal (or toroidal) 
direction~$\phi$. The problem has been considered both analytically 
(Aly 1984, 1991, 1993, 1995; Low 1986; van~Ballegooijen 1994; 
Lynden-Bell \& Boily 1994; Sturrock et al. 1995; Wolfson 1995;  
Uzdensky et al. 2002; Uzdensky 2002) and numerically (Barnes 
and Sturrock 1972; Yang et al. 1986; Porter et al. 1992; Wolfson 
\& Low 1992; Roumeliotis et al. 1994; Miki{\'c} and Linker 1994; 
Uzdensky et al. 2002).

\begin{figure}
\plotone{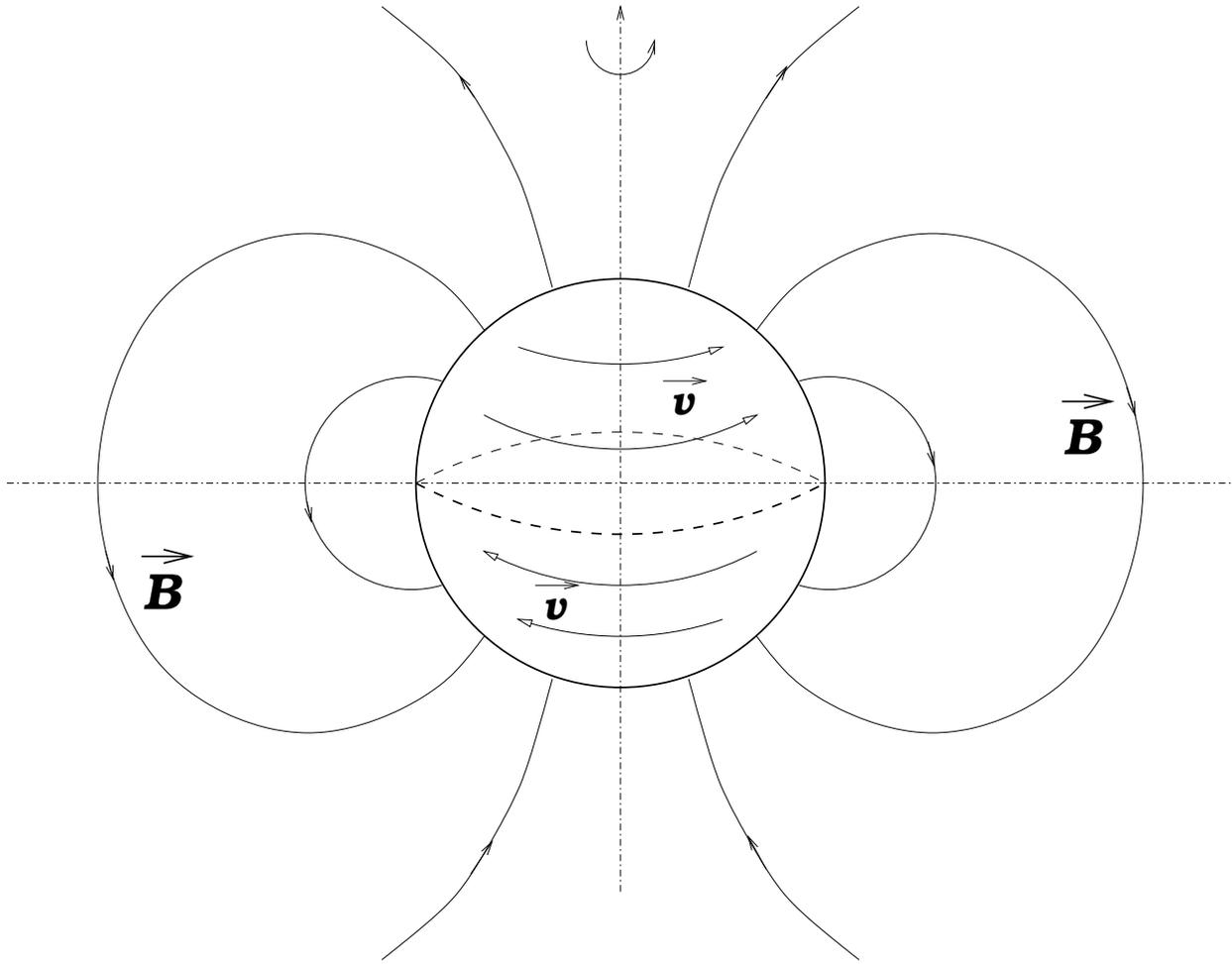}
\figcaption{Spherical geometry: axisymmetric dipole-like
field outside a sphere. The azimuthal displacement of the
footpoints on the sphere is represented by~${\bf v}$.
\label{fig-spherical}}
\end{figure}

\begin{figure}
\plotone{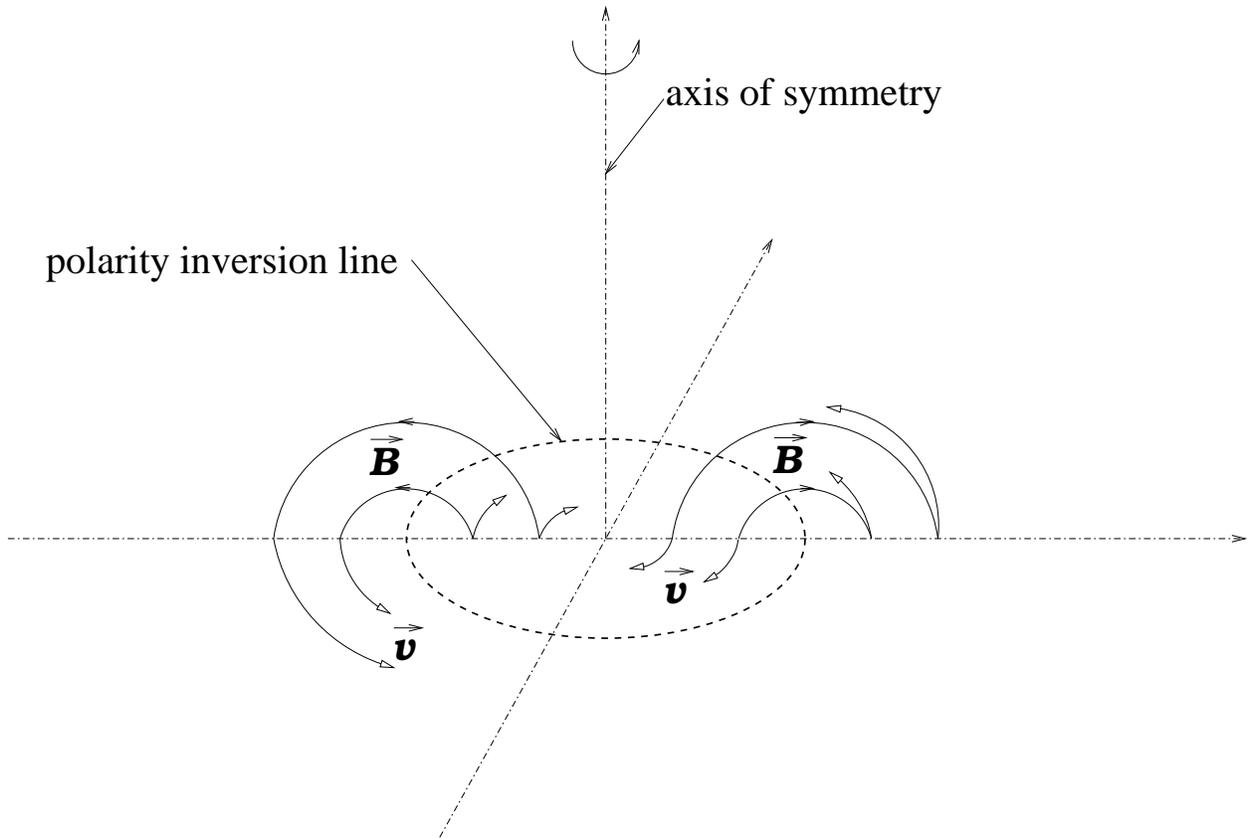}
\figcaption{Cylindrical geometry with axial symmetry around
the vertical axis. The magnetic field comes out of the plane 
$z=0$ inside the circular polarity inversion line and goes back 
into this plane outside this line. The azimuthal displacement of 
the footpoints on the plane is represented by~${\bf v}$.
\label{fig-cylindrical}}
\end{figure}

One should be aware that the are actually two distinct geometrical settings
in the axisymmetric case. One (which we shall call {\it spherical geometry}) 
is where the domain if interest is the outside of a differentially rotating 
sphere, with all the footpoints fixed on the surface of the sphere (studied, 
for example, by Low 1986; Wolfson \& Low 1992; Miki{\'c} \& Linker 1994; 
Roumeliotis et al. 1994; Wolfson 1995; Aly 1995; Sturrock et al. 1995). 
This case is usually considered in the context of solar corona. The field 
topology here is that of a point dipole placed inside the sphere (see 
Fig.~\ref{fig-spherical}). The other case, superficially similar to the 
Cartesian one, is {\it cylindrical geometry}, where the domain of interest 
is the half-space above an infinite plane on which all the footpoints are 
fixed (considered by Barnes and Sturrock 1972; Yang et al. 1986; Porter et 
al. 1992; Lynden-Bell \& Boily 1994; Sturrock et al. 1995 among others). 
The field topology in the cylindrical case can be visualized by, for 
example, placing a ring dipole on a plane surface or by putting a point 
dipole (with its axis being perpendicular to the plane) underneath the 
plane surface (see Fig.~\ref{fig-cylindrical}). This geometry is 
relevant to both the solar and accretion disk studies. Finally, 
a magnetically-linked star--disk system involves a combination 
of these two settings (van~Ballegooijen 1994; Goodson et al. 1999; 
Uzdensky et al. 2002). 


\section{Sequence of Force-Free Equilibria: Generating 
Function Method vs. Prescribed-Shear Approach}
\label{sec-2approaches}

In either geometry, a force-free field is described by the 
Grad--Shafranov equation. In Cartesian  geometry the equation 
is 
\begin{equation}
{{\partial^2\Psi}\over{\partial x^2}} + 
{{\partial^2\Psi}\over{\partial y^2}} = -F F'(\Psi)\, ,
\label{eq-GS-cartesian}
\end{equation}
and in the axisymmetric geometry (in spherical coordinates),
\begin{equation}
{{\partial^2\Psi}\over{\partial r^2}}+
{{\sin\theta}\over{r^2}} {\partial\over{\partial\theta}}
\left({1\over{\sin\theta}} {{\partial\Psi}\over{\partial\theta}}\right)=
-F F'(\Psi)\, .
\label{eq-GS-axisymmetric}
\end{equation}

Here, $\Psi$ is the magnetic flux function, related to magnetic 
field ${\bf B}$ via
\begin{eqnarray}
B_x &=& -{{\partial\Psi}\over{\partial y}}\, ,\\
B_y &=& {{\partial\Psi}\over{\partial x}}\, ,
\label{eq-Psi-cartesian}
\end{eqnarray}
in the Cartesian case, $\Psi=\Psi(x,y)$, and via
\begin{eqnarray}
B_r &=& {1\over{r^2\sin{\theta}}}\, 
{{\partial\Psi}\over{\partial\theta}}\, , \\
B_\theta &=& -{1\over{r\sin{\theta}}}\, 
{{\partial\Psi}\over{\partial r}}\, . 
\label{eq-Psi-axisymmetric}
\end{eqnarray}
in the axisymmetric case, $\Psi=\Psi(r,\theta)$.
Note that $\Psi$ has different dimensionality in the two cases, because in 
the Cartesian case it is defined as the magnetic flux per unit {\it length} 
in the $z$ direction, whereas in the axisymmetric case it is defined as the 
magnetic per unit {\it angle} in the azimuthal direction~$\phi$.

The function $F(\Psi)$ that appears on the right-hand side of 
equations~(\ref{eq-GS-cartesian}) and~(\ref{eq-GS-axisymmetric}) 
is called the {\it generating function} and is related to the 
magnetic field component in the ignorable direction:
\begin{equation}
F(\Psi) = B_z
\label{eq-F-cartesian}
\end{equation}
in the Cartesain case and
\begin{equation}
F(\Psi) = B_\phi r \sin{\theta}
\label{eq-F-axisymmetric}
\end{equation}
in the axisymmetric case.

Usually one considers equilibria with {\it closed} magnetic field
lines, where all the lines originate and terminate on the photosphere,
and none extend to infinity. The loss of equilibrium is then frequently
associated with the opening to infinity of at least a portion of the 
magnetic field.

Now, how should one pose a problem of calculating the sequence
of equilibria?

It is generally agreed that the proper way to set up the problem is 
to provide the boundary conditions on the photospheric surface by 
prescribing two functions (see, e.g., Priest \& Milne 1980; Klimchuk 
\& Sturrock 1989; Low 1990). The first is the magnetic flux distribution 
$\Psi_0$ on the bounding surface, which plays a role of a boundary 
condition for the flux function $\Psi$ in the Grad--Shafranov equation.
In Cartesian geometry this function is $\Psi_0(x)$ at $y=0$; in 
spherical geometry, where the domain of interest is the outside 
of a sphere, this function is $\Psi_0(\theta)$ on the sphere's surface, 
$r=R_*$; in cylindrical geometry this function is the flux distribution 
on the disk surface, i.e., $\Psi_0(r)$ at $\theta=\pi/2$. Finally, in 
the case of a magnetically-linked star--disk system, flux distributions 
on both the stellar surface, $\Psi(R_*,\theta)$, and the disk surface, 
$\Psi(r,\pi/2)$, need to be specified.

The most often considered (and the simplest) problem is the one in 
which the footpoints of the field lines move only in the ignorable 
direction, in which case the magnetic flux on the surface is fixed 
and we get time-independent boundary conditions. In this paper we 
are going to restrict ourselves to this case.

[Note that in some studies this could not be done, because these 
studies were done in a framework of a self-similar (in spherical 
radius $r$) model developed by Lynden-Bell \& Boily (1994) in 
cylindrical geometry and by Wolfson (1995) in spherical geometry 
and extended to Cartesian geometry by Aly (1994). 
Because of the assumed self-similarity,
these models had no characteristic radial length scale, and so 
one could not specify the boundary conditions at any particular 
radius (where the footpoints might be rooted); one only needed 
to give the ($r$-independent) angular boundary conditions. As a 
result, the footpoints were not fixed on any surface but were 
allowed, and actually had to, move in the non-ignorable direction 
(i.e., meridionally). This meridional motion of the footpoints was 
computed only a posteriori, together with the shearing motion in 
the ignorable direction.]

In the physically-motivated problem we are discussing here, 
the second function to be prescribed is the (time-dependent)
connectivity of the footpoints of the magnetic field lines. 
This means that for each field line~$\Psi$ one should specify 
the relative displacement (in the ignorable direction) between 
the line's two footpoints: $\Delta z (\Psi)$ in the Cartesian 
case or the twist%
\footnote
{We shall use the word ``twist'' throughout this paper, even though
the word ``writhe'' would be, perhaps, more appropriate.}
$\Delta\Phi(\Psi)$ in the axisymmetric case.
This relative displacement is related to the ignorable-direction
component of the field, and hence to the generating function $F(\Psi)$, 
via
\begin{equation}
\Delta z(\Psi) = F(\Psi) \int\limits_\Psi {{dl}\over{B_{\rm pol}}} = 
F(\Psi) \int\limits_\Psi {{dx}\over{B_x}}\, ,
\label{eq-shear}
\end{equation}
\begin{equation}
\Delta\Phi(\Psi) = F(\Psi) \int\limits_\Psi 
{{dl}\over{B_{\rm pol} r^2 \sin^2{\theta}}} = F(\Psi) \int\limits_\Psi 
{{d\theta}\over{B_\theta r(\Psi,\theta) \sin^2{\theta}}}\, . 
\label{eq-twist}
\end{equation}
where the integrals are taken along the field line from one footpoint 
to the other. Note that the field-line integral in the Cartesian case 
is just the area (in the $x$--$y$ plane) per unit of poloidal flux.

Thus we see that, in this proper problem setting, the generating 
function~$F$ is not explicitly given. Instead, it is the footpoint 
displacement $\Delta z(\Psi)$ [or~$\Delta\Phi(\Psi)$] that one has
to prescribe, while~$F$ is determined implicitly via equations~(\ref
{eq-shear}) and~(\ref{eq-twist}). This means that one needs to know 
the solution $\Psi(x,y)$ [or $\Psi(r,\theta)$] in order to calculate 
$F(\Psi)$ for given~$\Delta z(\Psi)$ [or~$\Delta\Phi(\Psi)$]. However, 
from the practical point of view of actually attempting to solve the 
problem, one can see that in fact it is the generating function 
$F(\Psi)$ that we need to know in order to solve the Grad--Shafranov 
equations~(\ref{eq-GS-cartesian}) and~(\ref{eq-GS-axisymmetric}). 
Therefore, in order to make the task easier, several authors employed 
the so-called {\it generating function method} (GFM) (e.g., Low 1977, 
1982, 1990; Jockers 1978; Birn et al. 1978; Priest \& Milne 1980). 
In this method, one actually explicitly specifies the generating 
function~$F(\Psi,t)$, along with the flux on the boundary. Then 
one solves the Grad--Shafranov equation, and only after that does 
one calculate~$\Delta z(\Psi)$ or~$\Delta\Phi(\Psi)$ using equation~(\ref
{eq-shear}) or equation~(\ref{eq-twist}). When doing so, one typically 
fixes the functional dependence of~$F(\Psi)$, i.e., one writes
\begin{equation}
F(\Psi,t)=\lambda(t) f(\Psi)\, ,
\label{eq-def-lambda}
\end{equation}
where $f(\Psi)$ is some prescribed function.%
\footnote
{An alternative realization of the GFM is when one prescribes~$F(\Psi,t)$ 
not in a simple separable way~(\ref{eq-def-lambda}), but rather in some 
other, less trivial manner, as, for example, it was done in the self-similar 
model considered by Lynden-Bell \& Boily (1994), Wolfson (1995), and Aly 
(1994). In this model, 
the self-similar power exponent~$p$ (where $\Psi\sim r^{-p}$ and $F(\Psi)=
\lambda\Psi^{1+1/p}$) changes with the sequence control parameter~$\lambda$, 
the dependence $p=p(\lambda)$ determined as a solution of an eigen-value 
problem. In the cylindrical and spherical cases Lynden-Bell \& Boily (1994) 
and Wolfson do find 
field opening and current-sheet formation at finite twist, while Aly (1994) 
finds that in the Cartesian case the opening occurs only asymptotically at 
an infinite shear. Even though both these findings are in agreement with 
the conclusions of the present paper, one must be cautious. For example, 
in Wolfson's (1995) spherical-case analysis the twist-angle profile, while 
remaining finite, developed a discontinuity across the equator 
as $p$ approached zero; thus, it is not clear whether the field 
opening at finite twist (and the associated current sheet formation) 
is related to the twist itself or to it's rapid change across
the polarity inversion line.
The reason why field-line opening is related to $p$ going to 
zero is elaborated upon by Uzdensky (2002).} 
Then one increases the overall magnitude $\lambda$ of~$F$
starting from zero (potential field). It is implied that 
increasing $\lambda$ is equivalent to increasing the field-line shear. 
For each $\lambda$ one solves for force-free configurations 
belonging to {\it the same topology class} as the initial 
potential field (one has to disregard solutions with different 
field topology, such as those with newly emerged magnetic islands, 
etc., because they are physically unaccessible to the system in ideal 
MHD --- see Low 1977). One finds then (Low 1977, 1982, 1986, 1990; 
Jockers 1978; Birn et al. 1978; Priest \& Milne 1980), that there 
is a limiting value of~$\lambda$, which we shall call~$\lambda_{\rm max}$, 
beyond which no force-free equilibrium can be found.%
\footnote
{The existance of an upper limit on $\lambda$ follows rigorously 
from an elegant virial theorem presented by Aly (1984) (see also
Low 1986). It also follows from some general mathematical theorems 
discussed, for example, by Birn et al. (1978).}
This limiting value often corresponds to a finite shear, 
so one might conclude that a loss of equilibrium occurs at this 
point, implying that any further shear increase will force the 
system into a dynamic, non-equilibrium phase. This interpretation 
was suggested (in Cartesian geometry) by Low (1977, 1982, 1990), 
Birn et al. (1978), and by Birn \& Schindler (1981) among others.

This point of view has been rightly criticized in the literature
(see, for example, Jockers 1978;  Priest \& Milne 1980; Aly 1984,
1985; Low 1986; Klimchuk \& Sturrock 1989) for the obvious reason 
that specifying $F(\Psi,t)$ and increasing $\lambda$ while keeping 
$f(\Psi)$ fixed does not constitute a physically relevant, valid 
thought experiment. This is because there is no one-to-one 
correspondence between~$F$ and~$\Delta z$ or~$\Delta\Phi$. 
As we discussed above, a valid thought experiment would be, 
for example, to specify the shape of the relative displacement 
function~$z(\Psi)$ [or the twist function~$\Delta\Phi(\Psi)$] 
and to gradually increase it's overall magnitude:
\begin{equation}
\Delta z(\Psi,t)=\mu(t) \xi(\Psi)\, ,
\label{}
\end{equation}
\begin{equation}
\Delta\Phi(\Psi,t)=\mu(t) \xi(\Psi)\, ,
\label{eq-def-mu-axisymmetric}
\end{equation}
where $\xi(\Psi)$ is a prescribed function, and $\mu$ is increased
starting from zero (potential field). 

Then, the above-described scenario based on GFM is replaced by
the following picture. The basic idea here is that, upon reaching 
the value of the shear parameter $\mu_1$ that (in some crude sense) 
corresponds to~$\lambda_{\rm max}$, a system subject to continuously 
increased shear will evolve smoothly past this point in such a manner 
that $\lambda$ will {\it decrease}. If this is the case, then a system 
evolving under a prescribed displacement of the footpoints does not 
exhibit a loss of equilibrium at~$\mu_1$ (or~$\lambda_{\rm max}$). 
It important to note that during this evolution not only the overall 
magnitude, but also the functional form of the function $F(\Psi)$ will 
change with~$\mu$. 

Several studies have been performed in order to determine whether 
such a scenario may be realized in practice. To do this one has to 
be able to compute a force-free equilibrium for a specified~$\Delta 
z(\Psi)$ [or~$\Delta\Phi(\Psi)$]. This direct approach to the problem 
is obviously more difficult than the GFM and evidently requires numerical 
tools. One can use a complicated iteration procedure such as those devised 
by Finn \& Chen (1990) and by Wolfson \& Verma (1991) for Cartesian geometry 
(in solar context) and by Uzdensky et al. (2002) for axisymmetric geometry 
(in the accretion disk context). However, the best way to approach this 
problem seems to be the {\it magneto-frictional method} (MFM). This method 
was first introduced by Yang~et~al. (1986) in cylindrical geometry, and 
then subsequently used by Porter~et~al. (1992) in the same geometry, by 
Roumeliotis~et~al. (1994) in spherical geometry, and by Klimchuk~et~al. 
(1988) and Klimchuk \& Sturrock (1989) in Cartesian (line dipole) geometry. 
One should note, however, that most of these MFM studies have suffered from 
the effects of the computation domain having a finite size, which has 
noticeably inhibited field-line inflation. This limitation has been 
circumvented by Roumeliotis~et~al. (1994) who have introduced a log-$r$ grid, 
thereby effectively moving the outer boundary to a very large distance.

One particularly interesting example of how an MFM study was
used to criticize the GFM-inspired claim for a loss of
equilibrium at $\lambda=\lambda_{\rm max}$, is the paper by
Klimchuk \& Sturrock (1989). They used the MFM to analyze 
the particular configuration considered by Low (1977) and they
were able to show that smooth solutions with simple topology 
exist even for values of shear $\mu$ in excess of $\mu_1\equiv 1$ 
that corresponded to  $\lambda=\lambda_{\rm max}$ in Low's analysis.
Klimchuk and Sturrock have concluded that the loss of equilibrium 
at~$\lambda_{\rm max}$ was an artifact of the GFM imposing non-physical 
restrictions on the magnetic field in an inappropriate thought experiment.%
\footnote{
One has to add a word of caution, however: in their analysis, the shear 
was parametrized by~$\mu$ but was not simply proportional to it: for 
any given field line, the shear first increased but then decreased 
with~$\mu$.}

However, these findings of smooth evolution past the finite-shear
point corresponding to~$\lambda_{\rm max}$ do not contradict the 
idea that a loss of equilibrium may still occur at a greater (but 
still finite!) value of~$\mu$. To find out whether this would 
happen, one needs to understand what the endpoint of the sequence
would mean for the field configuration and for the behavior of the 
generating function~$F(\Psi,\mu)$.

Based on the body of evidence accumulated so far in the previous
studies, we can offer the following description of the evolution 
through the sequence of equilibria when $\mu$ is taken to be the 
control parameter. We shall discuss the axisymmetric case 
first. As $\mu$ is increased starting from zero, the corresponding 
shearing leads to the production of the magnetic field component 
in the ignorable direction, i.e., the azimuthal field~$B_\phi$. 
The magnetic field pressure associated with this component leads 
to the inflation of the field lines. On any given field line~$\Psi$,
the generating function $F(\Psi,\mu)$ first grows monotonically with~$\mu$ 
[starting from $F(\Psi,\mu=0)=0$, which corresponds to the initial potential 
field]. However, upon reaching a certain finite value of $\mu=\mu_1(\Psi)$ 
(generally, each $\Psi$ has its own value of~$\mu_1$), $F(\Psi,\mu)$ 
reaches a maximum, and cannot be increased further ( Aly 1984; Low 1986;
van~Ballegooijen 1994; Uzdensky et al. 2002).
At the same time the solution itself does not exhibit any singularities 
and there is no loss of equilibrium at this point. As $\mu$ is increased
even further, the equilibrium sequence continues: the field lines become 
more and more inflated, the energy of the system continues to increase 
with increased~$\mu$, but $F(\Psi,\mu)$ now decreases! 
If one introduces, for the sake of this discussion, some characteristic 
measure~$\lambda$ of the magnitude of function~$F(\Psi)$, then one
can say that a subsequent increase in $\mu$ past $\mu_1$ leads to a 
decrease in~$\lambda$ and the system's evolution enters the second 
stage [or the second branch of~$\mu(\lambda)$].

This double-valuedness of $\mu(\lambda)$ is one of the main reasons 
why the generating function method may be misleading: when using it, 
one is in danger of missing the second branch~$\mu>\mu_1$. Thus, using 
the GFM requires extreme care. The double-valuedness of $\mu(\lambda)$ 
was recognized by van~Ballegooijen (1994), who essentially used GFM to 
analyze his self-similar model. 
In this model (describing a uniformly rotating disk magnetically 
connected to a central star in cylindrical geometry) the shapes
of both the twist function $\Delta\Phi(\Psi)$ and the generating
function $F(\Psi)$ were fixed power laws [$\Delta\Phi(\Psi)$ was
in fact constant], while only their magnitudes changed. Therefore,
once it was taken into account that for each value of $\lambda<
\lambda_{\rm max}$ there are two solutions with two different values
of~$\mu$, the GFM description of evolution in terms of~$\lambda$ was 
equivalent to the description in terms of $\mu\equiv\Delta\Phi$. 
This fact enabled van~Ballegooijen to calculate his equilibria using 
the GFM and then present the sequence of solutions in the appropriate 
manner in terms of $\Delta\Phi$-evolution. In fact, Uzdensky et al. 
(2002) managed to reformulate van~Ballegooijen's system of equations 
in terms of the twist angle~$\Delta\Phi$, and thus were able to calculate 
the sequence by directly prescribing $\mu$ instead of~$\lambda$. 
The results were identical to van~Ballegooijen's.

As the evolution makes the transition from the first to the 
second branch, the system's behavior gradually starts to change.
While during the first branch the magnetic field is inflated not 
very strongly, the solution on the second branch is characterized 
by rapidly accelerating inflation. Aly (1995) used a very general and 
rigorous mathematical argument to show that the inflation is at least 
exponential in spherical geometry (as was also found analytically by 
Sturrock et al. 1995 in both the spherical and cylindrical geometries) 
and suggested that it may in fact be explosive, with the opening of 
the poloidal field reached at a certain {\it finite} value~$\mu_c$. 
This latter suggestion is supported by semi-analytical self-similar 
studies by Lynden-Bell \& Boily (1994), van~Ballegooijen (1994), 
Uzdensky et al. (2002) (in cylindrical geometry), and by Wolfson (1995) 
(in spherical geometry). It is also supported by studies of numerically 
calculated sequences of force-free equilibria (Roumeliotis et al. 1994; 
Uzdensky et al. 2002), as well as by the full MHD numerical simulations 
by Miki{\'c} \& Linker (1994).%
\footnote
{Similar results were obtained in a 3D numerical simulation by 
Linker \& Miki{\'c} (1995) in the case of an axisymmetric partially 
open configuration and by Amari et al. (1996b) in the case of a single 
coronal flux tube having its footpoints twisted by slow photospheric 
motions.}
As $\mu\rightarrow \mu_c$, the field asymptotically approaches 
the open state, the magnetic field energy reaches its maximum 
and there are no equilibrium solutions beyond~$\mu_c$ that have 
the same magnetic field topology (closed field). One can say 
that at this point a loss of equilibrium has occurred [this 
situation corresponds to {\it Global Singular Nonequilibrium}
introduced by Aly (1993)].
Interestingly, Yang et al. (1986) and Porter et al. (1992), 
applying the MFM to the cylindrical case, have observed 
a drastic field-line expansion at finite twist angles, 
but have not claimed to have found any evidence for 
a finite-twist loss of equilibrium. This is probably 
explained by the fact that in their calculations all 
the field lines had to be confined inside a finite-size 
computation box with infinitely conducting outer boundaries, 
a limitation recognized and acknowledged by the authors.

As the system approaches the critical twist, i.e., as $\mu\rightarrow\mu_c$, 
the function $F(\Psi,\mu)$ approaches zero. This is is actually easy to 
understand: the asymptotic endpoint of the sequence is the open state, 
where the entire region is divided into two domains, separated by a 
current sheet. In one domain the magnetic field lines go from the surface 
to infinity, and in the other domain exactly the same number of field lines 
return from infinity to the surface. In the three-dimensional space the field 
is mostly radial and its magnitude drops off with radius~$r$ as~$r^{-2}$; the 
energy of the field is the upper bound of the energies of all the states with 
closed field lines and the same boundary conditions (Aly 1984, 1991; Sturrock 
1991). The open field in each domain is potential, hence $F(\Psi)=0$.

[It is actually more likely that one has a {\it partial field-line 
opening} where, in addition to two open-field regions, there 
is a region of closed field lines near the polarity-inversion
line (e.g., Low 1986; Wolfson \& Low 1992; Uzdensky 2002). In this 
case different field lines may have different critical (opening) 
value of~$\mu$: $\mu_c=\mu_c(\Psi)$. The smallest of all these 
values, $\mu_{c,\rm min}$, will mark the beginning of the partial 
field opening process; as $\mu$ is increased past $\mu_{c,\rm min}$,
more and more field lines will open (see Uzdensky 2002).]

Now let us consider the case of Cartesian geometry. 
Once again, one starts with the potential field, $\mu=0$,
$\lambda=0$, and gradually increases~$\mu$. The first stage 
of the evolution is similar to the axisymmetric case: the
field starts to expand gradually and $\lambda$ increases
[in this discussion $\lambda$ is again taken as some characteristic
measure of the magnitude of~$F(\Psi$)]. At some finite value 
$\mu=\mu_1$, $\lambda$ reaches a maximum $\lambda_{\rm max}$ 
and then starts to decrease.
This non-monotonic behavior of $\lambda(\mu)$ [and hence of the axial 
field $B_z\equiv F(\Psi,\mu)$] with increased shear in Cartesian geometry 
was found numerically by Jockers (1978) and, analytically, by Birn~et~al. 
(1978), whose explicit self-similar solutions were later generalized and
discussed by Priest \& Milne (1980). This behavior was also observed in 
numerically-constructed equilibrium sequences by Klimchuk et al. (1988) 
and by Finn \& Chen (1990) and, in full MHD simulations, by Biskamp 
\& Welter (1989) and by Amari et al. (1996a). What's important is that the 
system does not experience any loss of equilibrium at this point. In this 
respect the behavior in Cartesian geometry is identical to that in the 
axisymmetric case. However, at larger values of the shear parameter~$\mu$, 
as the system enters the second stage of the expansion process, the behavior
in the Cartesian case starts to deviate qualitatively from that in the 
axisymmetric case. The field lines still expand more and more in response 
to an increase in~$\mu$ and gradually approach the open state. But, what 
is most important, closed-field equilibrium configurations exist for {\it 
arbitrarily large} values of shear and no finite-time loss of equilibrium 
develops. The opening of the field lines (accompanied by $\lambda\rightarrow 
0$) is achieved only asymptotically in the limit $\mu\rightarrow \infty$. 
This fact has been shown by Aly (1985; see also Aly 1990, 1993) employing 
a general mathematical argument, and also demonstrated by Aly (1994) by 
applying Lynden-Bell\& Boily's (1994) self-similar model to plane geometry. 
The same conclusion is also supported by Birn~et~al. (1978), by MFM studies 
by Klimchuk~et~al. (1988), and by full MHD numerical simulations by 
Amari~et~al. (1996a).

It is interesting to note that the above two-branch scenario for 
the evolution in the Cartesian case is valid only in the case when 
the horizontal expansion of the field is not inhibited. If, however, 
the magnetic flux is confined in a finite-size box or between two 
vertical walls so that the field lines are not allowed to expand 
freely in the horizontal direction, then the may be no second branch. 
In this case, $\lambda$ will increase monotonically with increased shear 
and will asymptotically reach a maximum limiting value~$\lambda_{\rm max}$
as $\mu\rightarrow \infty$, while the field expands vertically and approaches
the open state (see an explicit const-$\alpha$ example by Priest \& 
Forbes 1990, numerically-computed sequences of equilibria by Finn \& 
Chen 1990 and by Wolfson \& Verma 1991, and MHD simulations by Biskamp
\& Welter 1989). 

Note also that the fact that $\lambda$ and hence $B_z$ are bounded
can also be used to construct a very elegant argument showing that
the opening of the field requires infinite footpoint displacement
in the Cartesian case (Low 1986; see also Low 1990; Aly 1993). Indeed, 
in Cartesian geometry the energy (per unit length in the~$z$ direction) 
of even a partially-open field is infinite. The energy input into the 
force-free field is due solely to the work done against the magnetic 
forces by the footpoint motions. Thus, the energy input rate per unit 
surface area is proportional to the product of the ignorable and normal 
to the surface components of the magnetic field, times the footpoint 
velocity. Since the normal field component stays fixed and the ignorable 
component is bounded, the total energy pumped into the corona is always 
finite (e.g., Aly 1985) and hence all the field lines are closed for any 
finite footpoint displacement.

To sum up, in the axisymmetric case we have a finite time 
(finite~$\mu$) field opening: force-free equilibria of closed magnetic 
field exist only in a finite range of twist angles ($\mu<\mu_c$).
We may say that a loss of equilibrium occurs at $\mu=\mu_c$.
On the other hand, in the case of Cartesian geometry the opening point 
$\mu_c$ is moved to infinity and loss of equilibrium occurs only at
infinite shear.


\section{Finite-Time Field Opening: Why Is It Expected 
in the Axisymmetric Geometry but Not in the Cartesian One?}
\label{sec-finite-time-opening}

In the rest of this paper we present very general and very simple 
scaling arguments explaining why finite time opening is expected to 
occur in the axisymmetric geometry setting, but not in the Cartesian 
one. Our arguments are based on a fundamental difference of the 
mathematical structure of the Grad--Shafranov equation in the two
cases and have a transparent physical (or, rather, geometrical) 
interpretation. Our intent here is to to bring out what we believe 
is the physical essense of the opening phenomenon, even at the
cost of sacrificing mathematical rigorousness for the sake of 
physical clarity. Thus, our arguments should be regarded as a 
supporting evidence --- not as a formal proof.

We consider a system of nested flux surfaces and we count the 
magnetic flux $\Psi$ from the outside inward, with the outermost
flux surface labeled by~$\Psi=0$. For simplicity, we first restrict 
our consideration to a situation where there are no thin structures 
formed in the corona.

Let us first consider the axisymmetric case (see also Uzdensky~et~al. 2002).
Consider the small-$F$ limit of equation~(\ref{eq-GS-axisymmetric}), in 
which the nonlinear term $FF'(\Psi)$ for some chosen field line~$\Psi$ 
is smaller than each of the two linear term on the left-hand side at 
typical distances of order the footpoint radius~$r_0(\Psi)$. 
By dimensional analysis, this implies
\begin{equation}
F(\Psi) \ll {\Delta\Psi\over{r_0(\Psi)}}\, ,
\label{eq-small-F-axisymmetric}
\end{equation}
where $\Delta\Psi$ is the amount of flux participating in the expansion
process.

Small values of $F$ may correspond to two qualitatively different solutions.
The first solution is close to the potential field, with the nonlinear
term being unimportant everywhere on the field line. The other
solution corresponds to greatly expanded field lines. In this 
solution, a given field line stretches out to distances so large
that the linear terms (dimensionally proportional to $\Delta\Psi/r^2$) 
become small and the nonlinear term gives an important contribution to 
the equilibrium structure of the distant portion of the expanded field
line. Then, for a given value of~$F$, we can define a characteristic 
radial scale~$r_F$ as 
\begin{equation}
r_F(\Psi,t) \equiv {\Delta\Psi\over{F(\Psi,t)}} \gg r_0(\Psi)\, .
\label{eq-def-r_F}
\end{equation}

The expression (\ref{eq-def-r_F}) also gives us an estimate
of the position~$r_{\rm ap}(\Psi,t)$ of the apex of the field line~$\Psi$ at 
the time when $F(\Psi)$ has the value~$F(\Psi,t)$. Hence we see that, as the 
field lines expand and approach the open state, $F(\Psi,t)\propto 
1/r_{\rm ap}(\Psi,t)$ goes to zero.

Our next step is to use this estimate to calculate the twist angle 
corresponding to the field line~$\Psi$ at time~$t$. This can be 
easily accomplished by using equation~(\ref{eq-twist}). One can 
write:
\begin{equation}
\Delta\Phi(\Psi,t) \sim
{{F(\Psi,t)}\over{(B_{\theta}r)_{\rm min}}}\, ,
\label{eq-DeltaPhi}
\end{equation}

Now, $(B_{\theta} r)_{\rm min} = 
-[(\partial\Psi/\partial r)/\sin{\theta}]_{\rm min}$ can be estimated 
simply as $\Delta\Psi/r_{\rm ap}(\Psi,t) \sim \Delta\Psi/r_F$. Thus, 
\begin{equation}
\Delta\Phi(\Psi,t)\sim F(\Psi,t){{r_F}\over{\Delta\Psi}} \sim O(1)\, .
\label{eq-twist-finite}
\end{equation}

Therefore, to lowest order in~$F$, the twist angle becomes 
independent of~$F$ as $F\rightarrow 0$ for a given field line. 
That is, as the field lines open up and $F\propto 1/r_{\rm ap}$
goes to zero, the twist angle~$\Delta\Phi$ approaches a finite 
value and we have a finite-time singularity. 

Now let us try to repeat the same scaling argument for the case of
Cartesian geometry. Let $\Delta x_0(\Psi)$ be the separation of the 
footpoints of a field line $\Psi$ in the $x$-direction, and let, at 
a given displacement $\Delta z(\Psi)=\mu\xi(\Psi)$, this field line 
rise above the plane to a maximum height~$y_{\rm ap}(\Psi,\mu)$.

Consider the small-$F$ limit: 
\begin{equation}
F \ll {{\Delta\Psi}\over{\Delta x_0^2}}\, .
\label{eq-small-F-cartesian}
\end{equation}
Once again, just as in the axisymmetric case, this limit is applicable 
in two cases: first, when the $z$-displacement is small, $\Delta z \ll 
\Delta x_0$, and the field is nearly potential, and second, when the 
displacement is large, and the field is far from potential and has 
expanded dramatically. In the latter situation the balance between 
the left-hand side and the right-hand side of equation~(\ref
{eq-GS-cartesian}) is achieved when the poloidal ($x,y$) field 
has expanded so drastically,
that its gradients are strongly reduced. In other words, the force-free 
balance requires that $B_{\rm pol}$ become small (compared with the field
near the surface $y=0$) along at least some portion of the field line, 
namely near the field line's apex. This in turn can only be achieved if 
the field has expanded strongly (in both the~$x$ and~$y$ directions) and 
thus approaches an open state. This balance introduces a new length-scale 
\begin{equation}
y_F\equiv {{\Delta\Psi}\over F} \gg \Delta x_0\, ,
\label{eq-def-y_F}
\end{equation}
which serves, just as in the axisymmetric case, as an estimate
for the distance $y_{\rm ap}(\Psi)$ from the $y=0$ plane to the 
apex of the field line, i.e., $y_{\rm ap}\sim y_F$.

Now let us estimate the displacement in the ignorable direction, 
$\Delta z$. Here, just as in the above argument for the axisymmetric 
case, we assume, once again, that no thin structures form in the corona, 
i.e., that uninhibited expansion takes place both in the~$y$ and~$x$ 
directions. Then the $x$ extent of an expanding field line, $\Delta x$,
will be of the same order as its $y$ extent, i.e., $\Delta x \sim y_{\rm 
ap} \sim y_F \gg \Delta x_0$ (e.g., Aly 1985; Finn \& Chen 1990).

The largest contribution to the integral~(\ref{eq-shear}) comes from 
the region of weakest poloidal field, i.e. the vicinity of the apex. 
Then we can estimate 
\begin{equation}
\Delta z \sim F {{y_F^2}\over{\Delta\Psi}} \sim y_{\rm ap} \sim
{{\Delta\Psi}\over{F}} \rightarrow \infty, 
\qquad {\rm as\ } F \rightarrow 0\, .
\label{eq-shear-infinite}
\end{equation}
Thus, there is no finite-shear field opening in this Cartesian-geometry
case. (A very similar argument for the Cartesian case was suggested by 
Finn and Chen 1990).

The difference in behavior between the axisymmetric and the Cartesian 
cases is clear and is of purely geometrical origin. In both cases, 
in order to maintain a strongly expanded flux tube in a force-free 
magnetostatic balance, the axial (azimuthal in the axisymmetric case) 
field should be of order of the poloidal field along a significant 
portion of the field line. Hence, the extent, or the reach, of the 
field line in the ignorable ($z$ or~$\phi$) direction (which can be 
estimated by looking at scales of order $r_{\rm ap}$ or $y_{\rm ap}$) 
should be of the same order as in the poloidal direction, i.e., 
as~$r_{\rm ap}$ or~$y_{\rm ap}$. For a strongly expanded field line, 
this distance scale is much larger than the initial footpoint separation 
($r_0$ or~$\Delta x_0$). In the Cartesian case this means $\Delta z\sim 
y_{\rm ap}\propto 1/F \rightarrow\infty$ as $F\rightarrow 0$, i.e., 
expansion to a large distance $y_{\rm ap}$ requires a shearing by 
an equally large distance~$\Delta z$. Hence, the end of the equilibrium 
sequence is only achieved in the limit of infinite shearing, and there 
is no finite-time field opening. In contrast, in the axisymmetric case, 
a large toroidal extent (of order $r_{\rm ap}\sim \Delta\Psi/F$) 
of the entire field line does not require a large displacement 
of the footpoints: a finite twist angle~$\Delta\Phi$, and hence 
a finite footpoint displacement (of order~$r_0\Delta\Phi$) is 
enough to produce an arbitrarily large length of the field line 
in the toroidal direction (of order~$\Delta\Phi r_{\rm ap}$).

Now, let us try to see whether the situation changes when one considers 
modifications due to additional geometrical constraints. In particular, 
in the above considerations we have assumed that there are no thin 
structures forming in the corona during the expansion process. Let 
us now see how the formation of such structures would change the 
system's behavior. We start again with the axisymmetric case.

In the axisymmetric case it is actually natural to expect (Aly 1991, 
1995; Lynden-Bell \& Boily 1994; Miki{\'c} \& Linker~1994; Wolfson~1995; 
Uzdensky~2002) that the expanding field lines will be strongly elongated 
along the line formed by the apexes of the field lines, i.e., along the 
so-called apex line, which we may approximate by a ray $\theta=\theta_{\rm 
ap}$. In the cylindrical case, $\theta_{\rm ap}$ is typically close 
to~$60^\circ$, while in the frequently considered spherical case with 
up-down symmetry the apex line coincides with the equator, i.e., 
$\theta_{\rm ap} = 90^\circ$. If one considers the region enveloped 
by a field line $\Psi$, then one finds that the $\theta$-extent of 
this region becomes much smaller than its radial extent, in other words, 
$\Delta\theta\ll 1$. In this case, the argument presented above will have 
to be modified somewhat, but, as we shall see, the main conclusion will 
remain the same.

Indeed, one can now employ a simple condition of magnetic pressure
balance across the apex line: the magnetic field at the apex is 
mostly toroidal, and, as one moves in the $\theta$-direction 
away from the apex at fixed~$r$, the field gradually becomes 
predominantly radial. The pressure balance then requires that 
$B_\phi[r_{\rm ap}(\Psi),\theta_{\rm ap}] \approx B_r^*(\Psi)$, 
where $B_r^*(\Psi)$ is the radial magnetic field just outside 
the apex region of field line $\Psi$, where the field is very 
close to the open potential field: 
$B_r^*(\Psi)=B_r[r=r_{\rm ap}(\Psi),\theta>\theta_{\rm ap}+\Delta\theta]$. 
Using equation~(\ref{eq-F-axisymmetric}), we then have
\begin{equation} 
F(\Psi) \approx B_r^*(\Psi)\, r_{\rm ap}(\Psi) \sin\theta_{\rm ap}\, . 
\label{eq-F-thin-axisymmetric}
\end{equation}

Equation~(\ref{eq-DeltaPhi}) remains essentially unchanged.
The only important difference comes from the fact that the 
toroidal field and hence the twist along the field line are 
concentrated near the apex in a rather narrow region of 
characteristic angular width~$\Delta\theta$. Therefore, 
the major contribution to the line integral on the right 
hand side of equation~(\ref{eq-twist}) comes from this 
region and we get
\begin{equation}
\Delta\Phi(\Psi,t) = F(\Psi,t) \int\limits_{\Psi}
{{d\theta}\over{B_{\theta} r(\Psi,\theta) \sin^2(\theta)}} \sim
F(\Psi,t) {\Delta\theta\over{(B_{\theta}r)_{\rm min}}}\, .
\label{eq-DeltaPhi-thin}
\end{equation}

Geometrically it is easy to see that for given $\Psi$,
$(B_\theta r)|_{\rm min}(\Psi) \simeq B_r^*(\Psi) 
\Delta\theta r_{\rm ap}$, and so 
\begin{equation}
\Delta\Phi(\Psi,t) \sim F(\Psi,t) {{\Delta\theta}
\over{ B_r^* \Delta\theta r_{\rm ap}}} = O(1)\, .
\label{eq-twist-finite-thin}
\end{equation}

Thus, we see that $\Delta\theta$ cancels out and we still get 
field line opening at a finite twist angle.%
\footnote
{A more elaborate treatment of thin structure formation in 
cylindrical geometry can be found in Uzdensky~(2002).}
 
Note that in these estimates the magnitude of $B_r^*$ did not matter, 
it actually dropped out of our equations. For the sake of completeness 
of our picture, however, we would like to give an estimate for~$B_r^*$. 
It can be estimated simply as the open field at radius~$r_{\rm ap}(\Psi)$;
in the case of freely expanding field lines, we have $B_r^*\simeq\Psi/r^2$, 
virtually independent of~$\Delta\theta$. If, however, the field lines are 
constrained to expand in a narrow conical region between two infinite radial 
walls separated by a poloidal angle $2\alpha\ll 1$, then $B_r^*\simeq\Psi/
\alpha r^2$. What is important, however, is that in either case $F(\Psi)
\sim \Psi/r_{\rm ap}(\Psi) \rightarrow 0$ as the field approaches open 
state (this, as we shall see, is dramatically different from the Cartesian 
case).

Consider now the Cartesian geometry in the case when the horizontal 
extent of an expanding field line, $\Delta x$, is much smaller than 
its vertical extent, $y_{\rm ap}$. Then the horizontal force-balance 
dictates that $F(\Psi)\equiv B_z$ be equal to the vertical field $B_y^*$ 
estimated outside of the apex region at the elevation~$y_{\rm ap}(\Psi)$: 
\begin{equation}
F(\Psi)=B_y^*(\Psi)\equiv B_y[|x|>\Delta x,y=y_{\rm ap}(\Psi)]\, .
\label{eq-F-thin-cartesian}
\end{equation}

The area-per-flux integral~(\ref{eq-shear}) can be roughly 
estimated as $y_{\rm ap}(\Psi)/B_y^*$, so
\begin{equation}
\Delta z(\Psi) \sim F {{y_{\rm ap}(\Psi)}\over{B_y^*}}
\sim y_{\rm ap}(\Psi) \rightarrow \infty \, .
\label{eq-shear-infinite-thin}
\end{equation}

Thus, we can see that in this case the opening of the field 
(i.e., $y_{\rm ap}\rightarrow \infty$) can be achieved only
asymptotically at an infinite footpoint displacement. 

Once again, one can notice that the actual value of $B_y^*$
drops out of the expression for~$\Delta z$. However, this 
value is very important if one wants to determine the time 
evolution of the function $F(\Psi,t)$ during the expansion. 
Here one can consider two possibilities. First, one can consider 
a case where a thin, current-sheet-like structure forms naturally 
during a horizontally unconstrained expansion (e.g., Aly 1985, 1994; 
Amari~et~al. 1996a). Then, the field outside the apex region can be 
estimated simply as $B_y^* \simeq \Psi/y_{\rm ap}$. Equation~(\ref
{eq-F-thin-cartesian}) then tells us that $F(\Psi,t)\propto 1/y_{\rm ap}
(\Psi,t)\sim 1/\Delta z(\Psi,t)\rightarrow 0$ as the field approaches the 
open state. This behavior is exactly the same as that in the Cartesian case 
without thin structures.

However, one can investigate another interesting situation, that 
of horizontally constrained expansion. Imagine that the entire flux 
system is confined between two vertical ideally conducting plates or 
walls (this configuration was considered by Biskamp \& Welter 1989; 
Priest \& Forbes 1990; Finn \& Chen 1990; and Wolfson \& Verma 1991). 
Let the separation $2a$ between the walls be of the same order as the 
typical footpoint separation~$\Delta x_0$. In such a system, the field 
lines can expand freely in the vertical direction, whereas their 
horizontal expansion is limited, $\Delta x <2a \sim \Delta x_0$. 
As the field lines expand vertically, the magnetic field outside 
of the apex region becomes close to a vertical potential field. 
Then $B_y^*$ can be estimated simply as~$\Psi/a$ [here we count 
the poloidal flux $\Psi$ from the wall: $\Psi(a)=\Psi(-a)=0$]. 
One can then immediately see that $F(\Psi)$ approaches a constant 
value as the expansion progresses:
\begin{equation}
F(\Psi,t) \sim {\Psi\over{a}} \rightarrow {\rm const}, \quad
{\rm as\ }  y_{\rm ap}\rightarrow \infty \, .
\label{eq-F-constrained-cartesian}
\end{equation}

We see that in the horizontally-constrained Cartesian case, the 
system's behavior changes dramatically in that important respect 
that $F$ does not decrease to zero, as in the unconstrained case, 
but instead reaches a finite value~$\Psi/a$. This behavior was in 
fact observed by Biskamp \& Welter (1989), Priest \& Forbes (1990), 
and by Wolfson \& Verma (1991); it was explained by Finn \& Chen 
(1990). One can describe this situation in terms of our parameters 
$\lambda$ and~$\mu$, by saying that $\lambda(\mu)$ is a monotonically 
increasing function, reaching $\lambda_{\rm max}$ only at $\mu=\infty$. 
Still we wish to emphasize that the main feature of the expansion process, 
namely the opening at infinite shear~$\Delta z$, stays unchanged.

One can easily generalize these results for the case of a flux system 
confined between two walls separated by a distance much larger than a 
typical footpoint separation~$\Delta x_0$, i.e., $a\gg \Delta x_0$. 
In this case, at first the system  expands freely both vertically and 
horizontally and~$F$ reaches a maximum value of order $F_{\rm max}\sim 
\Psi/\Delta x_0$ at $\Delta z = O(\Delta x_0)$. As the footpoints are 
displaced further, $F$ starts to decrease. When $\Delta z$ becomes 
comparable with~$a$, the system enters the horizontally-constrained
regime and $F(\Psi)$ asymptotically approaches a constant of order~$\Psi/a$.


\section{Discussion and Summary}
\label{sec-summary}

In this paper we investigated the question of finite-time opening 
of a force-free magnetic field evolving quasi-statically under slow 
footpoint motions on the boundary. This problem is of great importance
in studies of both the solar corona and accretion disk magnetospheres.
We started (in \S~\ref{sec-intro}) by discussing two principal classes 
of the problem's geometry that are most frequently considered: the 
Cartesian geometry with the translational symmetry along a straight 
line, and the axisymmetric geometry with the symmetry with respect 
to rotations around an axis. This latter class includes both the 
spherical geometry, where the domain of interest is the outside of 
a sphere, and the cylindrical geometry, where the domain of interest 
is the half-space above a plane. In \S~\ref{sec-2approaches} we 
introduced the set of equations governing the force-free evolution 
in these two geometrical settings. After that we gave a review of 
the existing literature focusing on various approaches and ways to 
describe the footpoint shearing responsible for driving the evolution. 
This followed by our compilation of the most commonly accepted (in our 
opinion) scenario for both the Cartesian and the axisymmetric cases. 
The main aspects of the evolution can be described as follows (see 
\S~\ref{sec-2approaches}).
In both cases the evolution consists of two phases. During the 
first phase the shape of the flux surfaces changes very little,
while the magnetic field's component in the ignorable direction 
(which in the simplistic setting considered here coincides with
the direction of the footpoint motion) grows monotonically in time. 
After some finite shearing, however, this ``toroidal'' component 
stops growing; the system enters the second phase, during which 
the ``poloidal'' (i.e., perpendicular to the ignorable direction)
field starts expanding rapidly, while the toroidal field on the
photospheric surface decreases. Eventually the system approaches 
the open (or partially-open) field configuration. The main difference
between the two types of geometry is that in the Cartesian case
the opening is achieved only asymptotically in the limit of infinite
shear, whereas in the axisymmetric case it is most likely achieved
at some finite shear (finite-time field opening). 

Finally, in \S~\ref{sec-finite-time-opening} we presented a series of
several simple physical arguments invoked to explain the geometrical
origin of this most important qualitative difference in the character
of the field-line expansion and opening process between the two cases. 
The basic idea of these arguments can be described as follows. The
field-line inflation is driven by the toroidal field's pressure,
which in a force-free equilibrium is balanced by the tension of 
the poloidal field. Hence, in the outer parts of a strongly-expanded
field line the toroidal field's strength should in some sense be 
comparable with that of the poloidal field. This, in turn requires 
that the toroidal extent of the expanded field lines be comparable 
with their very large poloidal extent. In the Cartesian case, this
condition automatically leads to the necessity of having a comparably
large footpoint displacement. Therefore, the opening of the field
cannot occur at any finite shear in this case. In the axisymmetric 
case, on the other hand, a very large toroidal extent of a field 
line's outermost portion can be reached at a finite footpoint 
rotation angle, which leads to a finite-twist field opening.

Now let us discuss some limitations of the simple picture described
in the present  paper.
One of the implied assumptions in the paper is that the opening 
is approached via a continuous sequence of stable equilibria.
In a very interesting alternative scenario presented (in spherical 
geometry) by Wolfson \& Low (1992), a sudden transition to a {\it 
partially-open state} is suggested to take place as soon as the
energy of the twisted but still closed magnetic field configuration
exceeds that of the partially open field (see also Low 1990).
Note that according to the Aly--Sturrock conjecture (Aly 1984, 1991; 
Sturrock 1991) the energy of the closed field can never exceed that 
of the {\it totally open} field. Therefore, only partial opening can
be achieved by such a sudden eruption. Also note that this scenario
can only work in the axisymmetric geometry because in the Cartesian
case the energy (per unit length in the ignorable direction) of even 
partially open field is infinite and hence can never be exceeded by
that of any closed-field configuration.

In addition, it is worth mentioning that the discussion in this paper
assumed that ideal MHD is valid throughout the entire evolution. If 
finite resistivity exists in the system, then reconnection may take 
place at a finite shear, leading to a change in the field topology, 
e.g., to the formation of a plasmoid (e.g., Amari~et~al. 1996a).
This reconnection process is likely to take place soon after the 
energy of the sheared arcade exceeds the energy of the complex-topology
configuration with the plasmoid (Aly 1990, 1991, 1993). This phase of 
the evolution may be very rapid and dynamic, perhaps characterized by 
the plasmoid ejection (Aly 1993; Amari~et~al. 1996a), hence providing 
a possible mechanism for coronal mass ejections.

Finally, it is important to acknowledge that this paper deals 
exclusively with the question of {\it existence} of equilibria; 
the very important issue of {\it stability} of these equilibria 
is completely left out. The main reason for this is that one 
cannot analyze the stability of solutions before establishing 
their existence and, mathematically, existence studies can be 
done completely independently of stability studies. From a 
practical perspective, however, it is clear that a quasi-static 
evolution is physically meaningful only if the sequence is made 
of continuous stable equilibria. It is possible, for example, 
that the first branch of the evolution is always stable, while 
the solutions on the second branch are unstable' the transition 
state between the two branches is then a marginally stable state. 
This is in fact the essence of Low's (1990) suggestion that the 
gradual quasi-static sequence of equilibria would end at this 
marginally stable bifurcation state. In particular, he argued 
that the presence of even a small plasma pressure may render 
this state unstable, leading to a subsequent violent phase 
of evolution. Thus, a stability analysis of strongly-inflated 
force-free equilibria is of crucial importance and definitely 
presents the next logical step in the study of sheared force-free
systems, but it falls outside of the scope of our present work.

I am grateful to J.~J.~Aly and B.C.~Low for their interesting 
and useful comments. I would like to acknowledge the support 
by the NSF grant NSF-PHY99-07949.


\section*{REFERENCES}
\parindent 0 pt

Aly, J.~J. 1984, ApJ, 283, 349

Aly, J.~J. 1985, A\&A, 143, 19

Aly, J.~J. 1990, Comput. Phys. Comm., 59, 13

Aly, J.~J. 1991, ApJ, 375, L61

Aly, J.~J. 1993, in Cosmical Magnetism, ed. Lynden-Bell
(Cambridge: Public. Inst. of Astronomy), 7

Aly, J.~J. 1994, A\&A, 288, 1012

Aly, J.~J. 1995, ApJ, 439, L63

Amari, T., Luciani, J.~F., Aly, J.~J., \& Tagger, M. 1996a,
A\&A, 306, 913

Amari, T., Luciani, J.~F., Aly, J.~J., \& Tagger, M. 1996b,
ApJ, 466, L39

Barnes, C.~W., \& Sturrock, P.~A. 1972, ApJ, 174, 659

Birn, J., Goldstein, H., \& Schindler, K. 1978, Sol. Phys., 57, 81

Birn, J., \& Schindler, K. 1981 In: Solar Flare Magnetohydrodynamics, 
ed. E.~R.~Priest (New York: Gordon \& Breach), 337 

Biskamp, D., \& Welter, H. 1989, Solar Phys., 120, 49

Finn, J.~M., \& Chen, J. 1990, ApJ, 349, 345

Goodson, A.~P., B{\"o}hm, K.-H., \& Winglee, R.~M. 1999, ApJ, 524, 142

Jockers, K. 1978, Sol. Phys., 56, 37

Klimchuk, J.~A., Sturrock, P.~A., \& Yang, W.-H. 1988, ApJ, 335, 456

Klimchuk, J.~A., \& Sturrock, P.~A. 1989, ApJ, 345, 1034

Linker, J.~A., \& Miki{\'c}, Z. 1995, ApJ, 438, L45

Lovelace, R.~V.~E., Romanova, M.~M., \& Bisnovatyi-Kogan, G.~S.
1995, MNRAS, 275, 244 

Low, B.~C. 1977, ApJ, 212, 234

Low, B.~C. 1982 Rev. Geophys. Space Phys., 20, 145

Low, B.~C. 1986, ApJ, 307, 205

Low, B.~C. 1990, ARA\&A, 28, 491

Lynden-Bell, D., \& Boily, C. 1994, MNRAS, 267, 146

Miki{\'c}, Z., \& Linker, J.~A. 1994, ApJ, 430, 898

Porter, L.~J., Klimchuk, J.~A., \& Sturrock, P.~A. 1992, ApJ, 385, 738

Priest, E.~R., \& Forbes, T.~G. 1990, Sol. Phys., 130, 399

Priest, E.~R., \& Milne, A.~M. 1980, Sol. Phys., 65, 315

Roumeliotis, G., Sturrock, P.~A., \& Antiochos S.~K. 1994, ApJ, 423, 847

Sturrock, P.~A., \& Woodbury, E.~T. 1967, Proc. Int.
School of Physics ``Enrico Fermi'', Course 39, 
``Plasma Astrophysics'' (New York: Academic Press), p. 155 

Sturrock, P.~A. 1991, ApJ, 380, 655

Sturrock, P.~A., Antiochos S.~K., \& Roumeliotis, G. 1995,
ApJ, 443, 804

Uzdensky, D.~A. 2002, ApJ, 572, 432

Uzdensky, D.~A., K{\"o}nigl, A., \& Litwin, C. 2002,
ApJ, 565, 1191

van~Ballegooijen, A.~A. 1994, Space Sci. Rev., 68, 299

Wolfson, R. 1995, ApJ, 443, 810

Wolfson, R., \& Low, B.~C. 1992, ApJ, 391, 353

Wolfson \& Verma, 1991, ApJ, 375, 254

Yang, W.-H., Sturrock, P.~A., \& Antiochos, S.~K. 1986, ApJ, 309, 383

\clearpage

\clearpage

\clearpage

\end{document}